\documentclass[10pt, aps,prx,superscriptaddress, twocolumn, floatfix,nolongbibliography]{revtex4-2}
\usepackage{graphicx}
\usepackage{xcolor}
\usepackage{textcomp}
\usepackage{siunitx}
\usepackage{bm}
\usepackage[normalem]{ulem}
\usepackage{amsmath, amssymb}
\usepackage{amsfonts}
\usepackage{adjustbox}
\usepackage[utf8]{inputenc}
\usepackage{placeins}

\newcommand{\blue}[1]{\textcolor{blue}{#1}}

\begin{document}

\title{\large{
Visualization of defect-induced interband proximity effect at the nanoscale
}}

\author{Thomas Gozlinski}
\thanks{These authors contributed equally to this work.}
\affiliation{Physikalisches~Institut,~Karlsruhe~Institute~of~Technology,~76131 Karlsruhe,~Germany}

\author{Qili Li}
\thanks{These authors contributed equally to this work.}
\email{Corresponding author: qili.li@kit.edu}
\affiliation{Physikalisches~Institut,~Karlsruhe~Institute~of~Technology,~76131 Karlsruhe,~Germany}

\author{Rolf Heid}
\affiliation{Institute~for~Quantum~Materials~and~Technologies,~Karlsruhe~Institute~of~Technology,~76131~Karlsruhe,~Germany}

\author{Oleg Kurnosikov}
\affiliation{Universite de Lorraine, Institute Jean Lamour, France}

\author{Alexander Haas}
\affiliation{Physikalisches~Institut,~Karlsruhe~Institute~of~Technology,~76131 Karlsruhe,~Germany}

\author{Ryohei Nemoto}
\affiliation{Department of Materials Science, Chiba University, 1-33 Yayoi-cho, Inage-ku, Chiba 263-8522, Japan}

\author{Toyo Kazu Yamada}
\affiliation{Department of Materials Science, Chiba University, 1-33 Yayoi-cho, Inage-ku, Chiba 263-8522, Japan}
\affiliation{Molecular Chirality Research Centre, Chiba University,1-33 Yayoi-cho, Inage-ku, Chiba 263-8522, Japan}

\author{Jörg Schmalian}
\affiliation{Institute~for~Quantum~Materials~and~Technologies,~Karlsruhe~Institute~of~Technology,~76344~Eggenstein-Leopoldshafen,~Germany}
\affiliation{Institut~für~Theorie~der~Kondensierten~Materie, Karlsruhe~Institute~of~Technology, 76131 Karlsruhe, Germany}

\author{Wulf Wulfhekel}
\affiliation{Physikalisches~Institut,~Karlsruhe~Institute~of~Technology,~76131 Karlsruhe,~Germany}
\affiliation{Institute~for~Quantum~Materials~and~Technologies,~Karlsruhe~Institute~of~Technology,~76344~Eggenstein-Leopoldshafen,~Germany}

\date{\today}

\begin{abstract}
The vast majority of superconductors have more than one Fermi surface, on which the electrons pair below the critical temperature $T_C$, yet their superconducting behavior can be well described by a single-band Bardeen-Cooper-Schrieffer theory. This is mostly due to interband scattering, especially in superconductors in the dirty limit, rigidly linking the pairing amplitude of the different bands. This effect has severely limited experimental studies of the complex physics of multiband superconductivity. In this study, we utilize the fact that elementary Pb - as a clean limit system - has two Fermi surfaces that are only weakly coupled by interband scattering, allowing the formation of two separate condensates. By studying crystallographic defects in the form of stacking fault tetrahedra with our millikelvin scanning tunneling microscope, we show how to locally tune interband coupling ranging from weak to strong coupling and modify the superconducting order parameters from two well separated gaps to one merged gap around defects. The experiments critically test the theory of multiband superconductors and give a route to access a wide range of predicted quantum effects in these systems.
\end{abstract}

\maketitle

\section{Introduction}
Microscopic attractive interactions between electrons that lead to conventional superconductivity are captured by the Bardeen-Cooper-Schrieffer (BCS) theory \cite{BCS1957Microscopic} in an elegant and transparent form. BCS theory successfully predicts the temperature dependence of the superconducting gap or the shape of the quasiparticle spectrum, as measured in a tunneling experiment. In some sense, the theory performs even better than expected. Although, in its most elementary form, it only considers a single electronic band and a perfectly clean environment, it can still describe the tunneling spectra measured on many real superconducting materials. As an extension, the finite effective quasiparticle lifetimes, e.g. due to scattering by non-magnetic crystal imperfections or inelastic scattering processes were also considered by Dynes and others \cite{dynes_direct_1978, herman_microscopic_2016}. 
Most superconductors in nature have more than one electronic band crossing the Fermi energy, which should allow electrons to scatter between different bands in the normal state and, in principle, also lead to superconductivity in multiple bands. Multiband superconductivity was first suggested by Suhl, Matthias and Walker (SMW) \cite{Suhl1959}, only two years after the seminal theory by BCS, and highlighted the case of two non-interacting bands. Without interband pairing interaction, the individual bands are expected to develop distinct superconducting gaps that emerge at distinct transition temperatures ($T_c$) \cite{Suhl1959}. The reason behind the single-band BCS theory's long success lies in the fact that scattering between multiple bands couples these distinct pairing amplitudes which now emerge at a single $T_c$. In the limit of strong coupling between bands of similar density of states at the Fermi level, the gap values merge to a common value due to an interband proximity effect \cite{McMillan1968, Noce1989} in momentum space \cite{Sung-Wong1967, Schopohl-Scarnberg1977}. This process can be due to interband electron-phonon or electron-electron interactions in clean systems.
It becomes unavoidable in 'dirty' superconductors or at temperatures close to $T_c$, where electron lifetimes become shorter. This leads to the appearance of a single pairing condensate with essentially a single lifetime-broadened superconducting gap \cite{Suhl1959, Sung-Wong1967, Tang1970, Chow1971}, that disguises the multiband nature of the material. According to Anderson's theorem, the gap and the transition temperature are robust against nonmagnetic scatterings for single-band conventional superconductors \cite{Anderson_theorem}. However, in full agreement with our above statements, generalizations to multiband superconductors show that nonmagnetic scattering changes the superconducting gap values as long as their magnitudes are distinct \cite{Schmalian2015}. Robustness against nonmagnetic impurities reemerges once a common gap value has been established.

With advances in the growth of high-quality superconductor crystals and improvements in low-temperature measurement techniques, it has become possible to discern the energy differences among individual superconducting gaps in several clean multiband systems \cite{Binnig1980, Nagamatsu2001, Noat2010signatures, Zehetmayer2010, ruby_experimental_2015, Du2016Scrutinizing}. Among these materials, bulk lead (Pb) stands out as an elemental two-band superconductor \cite{Floris2007, ruby_experimental_2015, Saunderson2020, Gozlinski2023SciAdv} that is available in exceptional purity and crystallinity. Pb displays two Fermi surfaces, a compact one and an open one (see e.g. \cite{Choy_2000} and Fig. \ref{fig:SFT1}(d) and (c)). Recent experimental results further indicate that the interband coupling in Pb is rather weak \cite{Gozlinski2023SciAdv}, such that at low temperatures, two superconducting gaps in the tunneling spectrum appear. 
The compact Fermi surface of Pb gives rise to the larger gap ($\Delta_2$) and the open Fermi surface to the smaller one ($\Delta_1$) \cite{Saunderson2020,Gozlinski2023SciAdv}.
This makes Pb an ideal model system for experimentally studying how two bands and their condensates interact, e.g. by studying how individual crystal defects alter the scattering within and between bands and change the pairing state. This allows direct insight into the microscopic scattering events that eventually limit many superconducting devices through the critical current and critical field \cite{mattis_theory_1958, abrikosov_theory_1959, bardeen_critical_1962, maki_persistent_1963}. Gaining control over the interband coupling and scattering processes in a two-band superconductor may even give access to a wide range of predicted quantum effects, such as solitons \cite{Tanaka2002, Yerin2011}, vortices with fractional flux \cite{Babaev2002}, non-Abrikosov vortices \cite{Cho2006}, topological knots \cite{Cho2006}, 
or the Leggett mode that describes excitations of the relative phase of weakly coupled bands \cite{leggett_number-phase_1966}.

\section{Results}
\subsection{Local superconducting density of states}
A Pb(111) single crystal was cleaned by several cycles of sputtering and annealing, resulting in a surface with monoatomic steps. It was then placed on our precooling station at $\sim \SI{78}{\kelvin}$, where it was rapidly cooled before insertion into the scanning tunneling microscopy (STM) head.
As a result of this procedure, few defects, which are typically challenging to prevent in the material, persist near the surface. In the constant-current STM topography Fig. 1(a), we can see three types of defects \footnote{A fourth type of defect are the small protrusions scattered throughout the image. These, however, are not intrinsic to the material or the preparation but caused by the interaction between surface and tip: In this case, an earlier (large) scale image was recorded, where the tip speed and feedback conditions were chosen in a way to minimize the time needed to find these SFTs, at the cost of occasional, unwanted tip-sample interactions at step edges, where single lead atoms would be pulled out and dragged along the line of scan before being dropped again.}: (i) small 3-5\,nm sized depressions of hexagonal shape that are well known and correspond to argon bubbles or small vacancy conglomerates near the surface \cite{muller_lateral_2016, muller_open-boundary_2017, song_observation_2017}; (ii) a screw dislocation, of which both ends exit the surface, leading to seemingly disrupted step edges and continuous connection of neighboring atomic terraces \cite{Wolf1991, Jakob2006, Aladyshkin2021}; (iii) a buried stacking fault tetrahedron (SFT) of about 70\,nm lateral width, which binds the screw dislocation \cite{Wolf1991} and is responsible for the weak interference pattern visible in Fig. 1(a). The SFT is an elementary crystal defect in fcc metals of low stacking fault energy \cite{silcox_direct_1959, Cotterill_1961, Yokota_1967, Bonsignori_1968, Wolf1991, Guan_stacking_2004, schaublin_irradiation-induced_2005, Hardy_2006, Matsukawa_2008, schouteden_lateral_2012, Wang_atomic_2013, Yu_removal_2013, schouteden_electronically_2016}. Like the smaller defects (i), it evolves from an agglomeration of vacancies that is too large in size to be energetically favoured \cite{uberuaga_direct_2007, schaublin_irradiation-induced_2005, zhang_radiation_2018}. Following topological rules on dislocation interactions, tensile stress is released in a process in which atoms rearrange and form a tetrahedral nanocrystal inside the bulk terminated by stacking faults \cite{silcox_direct_1959}. Its edges feature frustrated stair-rod dislocations and each face is oriented in a (111) direction of the crystal with an intrinsic stacking fault \cite{Ohmori1999, Osetsky2006}. Two common ways to reliably induce SFTs in a metal are quenching \cite{silcox_direct_1959} and ion irradiation \cite{schaublin_irradiation-induced_2005}. As we find a higher density of SFTs for rapid cooling than without, we argue that both argon ion sputtering and the quick cooldown are responsible for their formation in our case. Please see Appendices B for an SFT on the surface. A characteristic of SFTs near the surface are quantum well states (QWS) \cite{Guo_superconductivity_2004, Schackert_local_2015} that form between its triangular top (111) plane and the (111) sample surface (see Supplemental Material Note I). It is well established that these QWS allow to image sub-surface defects, including SFTs, with STM and to determine their depths below the surface \cite{kurnosikov_probing_2008, kurnosikov_long-range_2009, kurnosikov_internal_2011, 
 weismann_seeing_2009, schouteden_lateral_2012, schouteden_electronically_2016, muller_lateral_2016, muller_open-boundary_2017}. A sketch of the measurement principle is shown in Fig. 1(b).

\begin{figure*}[!htb]
    \centering
    \includegraphics[width=1\textwidth]{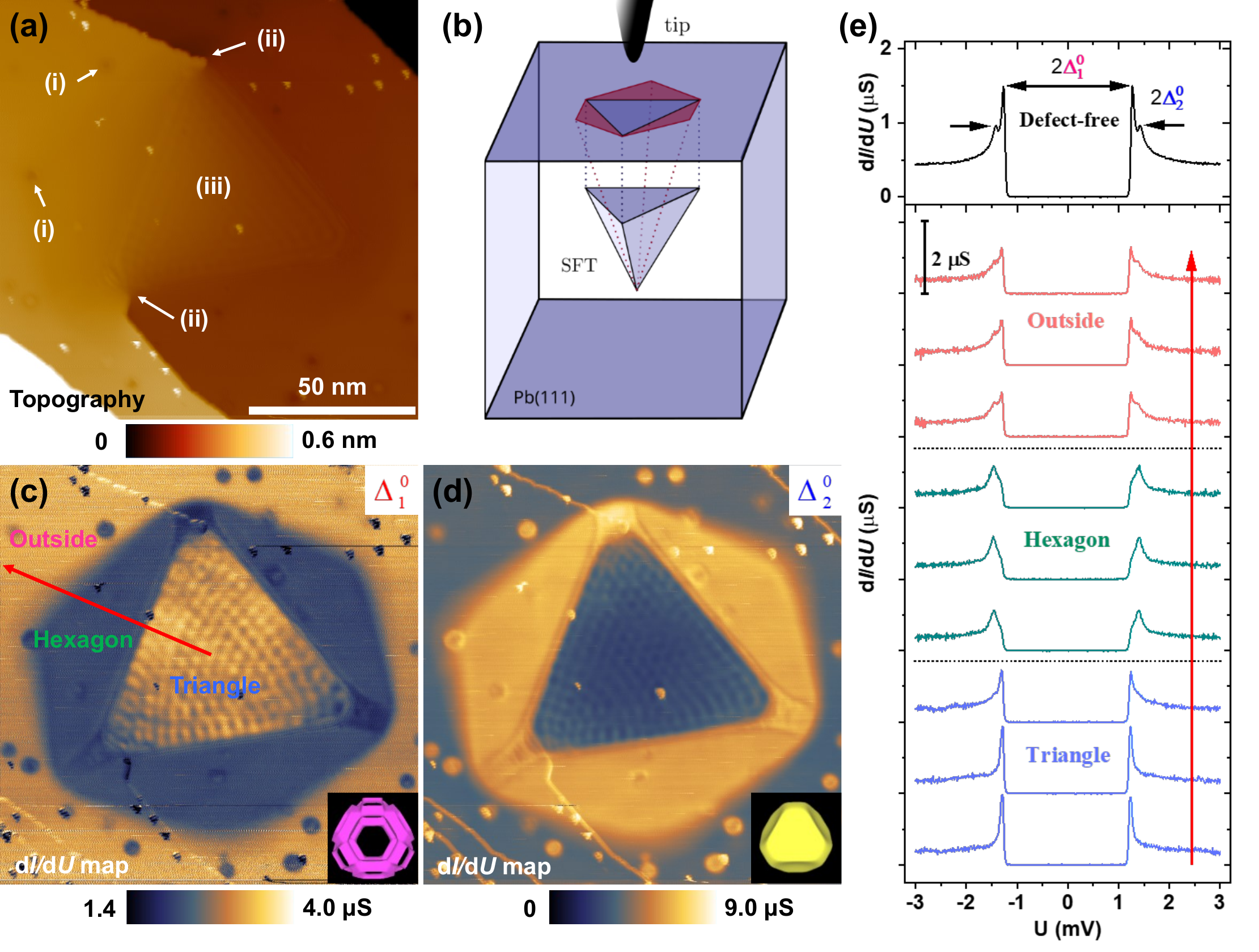}
    \caption{\textbf{Influence of stacking fault tetrahedron on superconducting gaps (SFT \#1).} (a) Topography measured with sample bias $U=1.4$ \si{\milli\volt} and tunneling current $I=1$ \si{\nano\ampere}. Three types of defects are indicated as (i) argon bubbles or small vacancy conglomerates, (ii) a screw dislocation and (iii) a buried stacking fault tetrahedron. (b) Sketch of the STM measuring a subsurface stacking fault tetrahedron in Pb(111). (c) d$I$/d$U$ mapping at $\Delta_1^0$. The inset shows the corresponding open Fermi surface \cite{Choy_2000}. (d) d$I$/d$U$ mapping at $\Delta_2^0$. The inset shows the corresponding compact Fermi surface \cite{Choy_2000}. The d$I$/d$U$ maps in (c) and (d) were measured with tip stabilized at $U=1.4$ mV and $I=1$ nA and lock-in modulation amplitude of $U_{mod}=50$ $\mu$V and frequency of 3.751 kHz. (e) d$I$/d$U$ spectra on defect-free area (top panel) and at different regions from triangle center to outside the SFT along the red arrow in panel (c). For d$I$/d$U$ spectra measurements across the SFT, the tip was stabilized at $U=3$ mV and $I=1$ nA, with lock-in modulation $U_{mod}=20$ $\mu$V for the amplitude and frequency of 3.751 kHz. For d$I$/d$U$ spectra on defect-free area, the tip was stabilized at the same tunneling condition with lock-in modulation amplitude of $U_{mod}=20$ $\mu$V and frequency of 3.041 kHz. }
    \label{fig:SFT1}
\end{figure*}

Taking spectra along the red line indicated in Fig. \ref{fig:SFT1}(c) confirms that the contrast in the d$I$/d$U$-maps in Fig. \ref{fig:SFT1}(c,d) stems from changes in the quasiparticle spectrum of the superconductor, i.e. the intensity of the coherence peaks and their energy [Fig. 1(e)].

In a two-band superconductor, the observed behavior can be either due to changes in the STM tunneling matrix element or a local change in the pairing condensate. Let us first comment why we believe that the former is not at play: The momentum dependence of the tunneling matrix element between the sample and the tip leads to a variation of the tunneling spectra when varying the crystal orientation \cite{tersoff_theory_1985,Noat2010signatures,ruby_experimental_2015}. We argue that this effect is absent in our case, as the probed surface is of (111) orientation throughout the sample.
Moreover, we studied more than 20 SFTs that all showed a common behavior namely [Fig. \ref{fig:different interband coupling}(d)-(f)]: 1. a higher coherence peak at $\Delta_2^0$ than at $\Delta_1^0$ in the hexagonal region irrespective of the depth and size of SFTs, 2. variations in both coherence peak intensity and energy in the triangular area for different SFTs. 
In the two distinct regions (triangle, hexagon) we see a variation in coherence peak energies, i.e. $1-(\Delta_{2}-\Delta_{1})/(\Delta_{2}^0-\Delta_{1}^0)$ up to $13\%$ for SFT \#2 [Fig. \ref{fig:different interband coupling}(e)] and almost merged coherence peaks for SFT \#3 [Fig. \ref{fig:different interband coupling}(f)] that cannot be explained by band sensitivity of tunneling, alone. 
The complete disappearance of the outer coherence peak inside the triangle (see SFT \#1 in Fig. \ref{fig:different interband coupling}(d)) also speaks against the former scenario. 
This lends credence to the other possibility, a variation in the coherence peak energies because of a local change in the two-band superconducting condensate. This scenario allows for both, intensity changes and energy variations in the coherence peaks.

\subsection{Influence of scattering on the two order parameters}
To understand our experimental tunneling spectra of the two-band superconductor Pb(111), we performed simulations of the density of states of an impurity-coupled two-band superconductor, following Ref. \cite{Sung-Wong1967}.
McMillan developed a model for superconductor-normal-metal sandwiches describing the proximity effect in the normal metal by scattering between the superconductor and normal metal bands \cite{McMillan1968}. Sung and Wong developed a model (S\&W model) describing a mutual proximity effect in a two-band superconductor induced by interband scattering at non-magnetic impurities \cite{Sung-Wong1967} which also includes intraband scattering. Both models result in coupled equations of the two order parameters. Finally, Schopohl and Scharnberg used the S\&W model in a particular form to model the DOS of two-band superconductors \cite{Schopohl-Scarnberg1977}. This model was shown to describe the tunneling spectra for MgB$_2$ \cite{Schmidt2002Evidence, Iavarone2002Two-band, Ekino2003Tunneling, Schmidt2003Break-junction, Silva2015Tunneling}, NbSe$_2$ \cite{Noat2010signatures, noat_quasiparticle_2015, Senkpiel2019Robustness}, Ba$_8$Si$_{46}$ \cite{Noat2010TwoEnergyGaps}, and RuB$_2$ \cite{Datta2020Spectroscopic}. 
It incorporates coupling between individual bands by allowing elastic interband scattering events with rates $\Gamma_{ij}$. This leads to a mutual proximity effect in $k$-space that couples the two condensates. As a consequence, the individual gap functions $\Delta_{1}$ and $\Delta_2$ become complex and energy-dependent. For generality, here we opt for the more general Sung and Wong model- also accounting for intraband scattering with rates $\Gamma_i$, as given in the Method section (see Appendices C for influence of intraband couplings and Supplemental Material Note IV for validity of the S\&W model).

\begin{figure*}[!htb]
    \centering
    \includegraphics[width=1\textwidth]{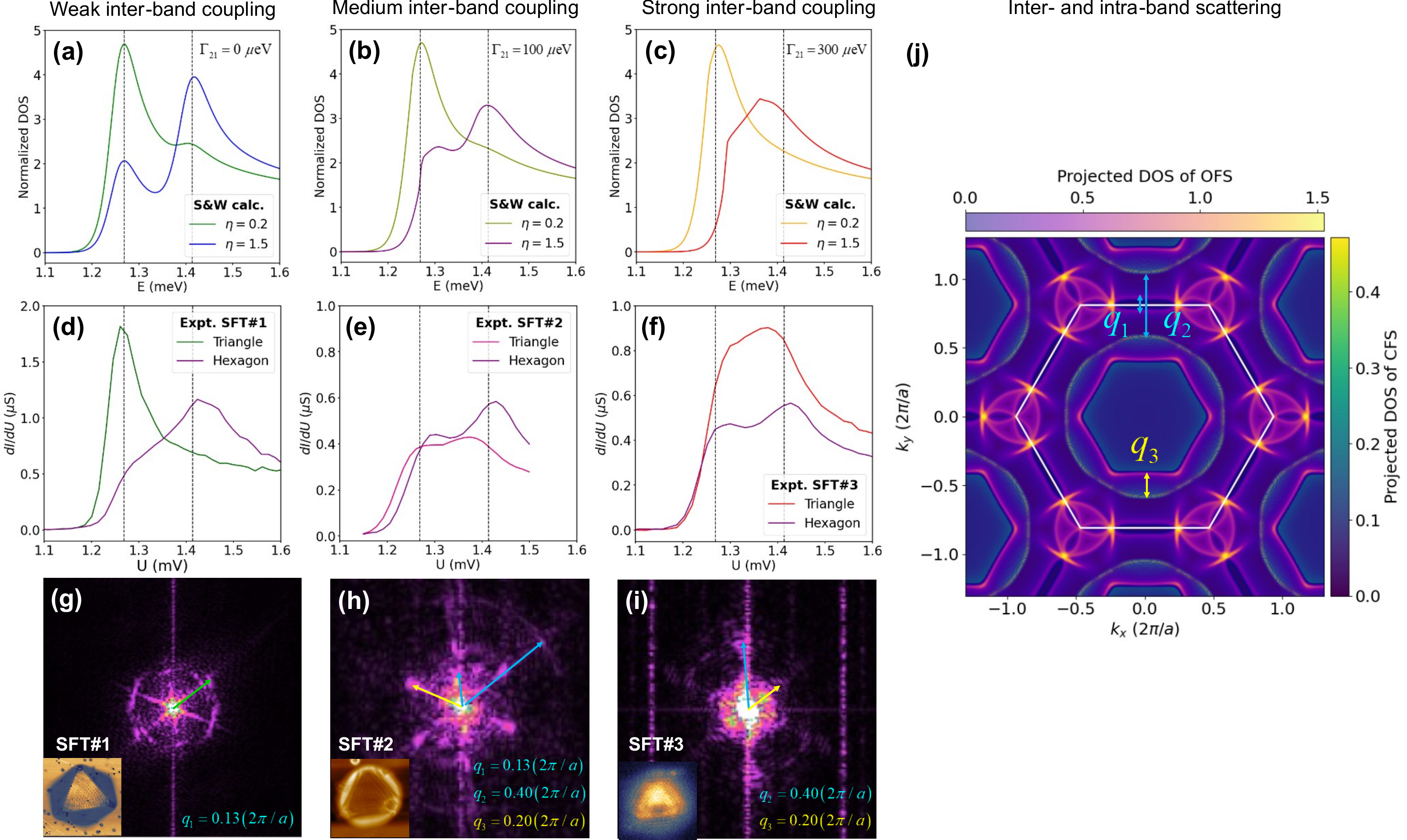}
    \caption{\textbf{Influence of scattering on two-band superconducting gaps.} (a)-(c) Demonstration of interband coupling on two-band superconducting gaps with different DOS ratio $\eta$  ($=N_2(E_F)/N_1(E_F)=\Gamma_{12}/\Gamma_{21}$) from weak to strong interband coupling. (a) No interband coupling. The intensity of coherence peaks depends on the DOS ratio of two bands, but without energy variation of coherence peaks. (b) Medium interband coupling. (c) Strong interband coupling. Only one coherence peak can be seen for the strong interband coupling. The green to orange color corresponds to increasing interband coupling for low DOS ratio. The blue to red color corresponds to increasing interband coupling for high DOS ratio. (d)-(f) Spectra at triangle and hexagonal regions for different SFTs. (d) Spectra for SFT \#1 (see also Fig. \ref{fig:SFT1}(e)). (e) Spectra for SFT \#2. (f) Spectra for SFT \#3.  Hexagonal regions always have a medium interband coupling with high DOS ratio, while the interband coupling and DOS ratio at triangular regions vary from weak to strong. The dashed lines are the energy of coherence peaks of defect-free spectroscopy in Fig. 1(e). In panels (e) and (f), the tip was stabilized at $U=100$ mV and $I=1$ nA, with tip height z offset -120 pm for the d$I$/d$U$ spectroscopy. The lock-in amplifier has a modulation with amplitude of $U_{mod}=20$ $\mu$V and frequency of 3.402 kHz. (g)-(i) QPI patterns. The insets are the corresponding d$I$/d$U$ maps that measured at $\Delta_1^0$ (SFT \#1), $\Delta_2^0$ (SFT \#2) and $\Delta_2^0$ (SFT \#3), respectively. (j) The calculated projected DOS of open Fermi surface (OFS) and compact Fermi surface (CFS). $k_x$ is in $[1\overline{1}0]$ direction, and $k_y$ is in $[11\overline{2}]$ direction. $k_z$ is in $[111]$ direction. The green arrows are the nesting vectors of intraband scattering. The yellow arrow is the nesting vector of interband scattering.
    }
    \label{fig:different interband coupling}
\end{figure*}

To illustrate the effects of interband coupling for Pb, we show model DOS calculations in Figure \ref{fig:different interband coupling} (a)-(c) with varying interband scattering, where the intrinsic gaps ($\Delta_1^0=1.252$ meV and $\Delta_2^0=1.40$ meV) and effective temperature ($T=139$ mK) were obtained from defect-free areas by fitting using the self-consistent equations in the S\&W model (see Appendices A Methods for details and Supplemental Material Note II).  The three panels display the self-consistently calculated normalized single particle DOS with different interband couplings. We show results for two different values of the ratio $\eta=N_2(E_F)/N_1(E_F)$ of the DOS of the two bands: $\eta=0.2$ and $\eta=1.5$. Naturally, for larger $\eta$ the higher energy coherence peak is more pronounced. In the case of no interband coupling $\Gamma_{21}=0$, the two gaps are fully decoupled, and the total spectrum is just the sum of two BCS gap functions [Fig. \ref{fig:different interband coupling}(a)]. With increasing $\Gamma_{21}$, the two gaps begin to merge [Fig. \ref{fig:different interband coupling}(b)] and eventually become indistinguishable [Fig. \ref{fig:different interband coupling}(c)].

With the above illustration, we are able to explain our experimental tunneling spectra for different SFTs. The spectra of three representative SFTs are shown in Fig.~\ref{fig:different interband coupling}(d)-(f). In the hexagonal region, c.f. purple curves in Fig. \ref{fig:different interband coupling}(d)-(f), an intermediate coupling brings the two gaps closer. In addition, the large DOS ratio 
$\eta$ makes the coherence peak intensity of $\Delta_2$ higher than that of $\Delta_1$. However, in the triangular region, the interband coupling and DOS ratio $\eta$ vary from weak (green curve in Fig. \ref{fig:different interband coupling}(d)) to strong (red curve in Fig. \ref{fig:different interband coupling}(f)). This is due to the QWS confined above the SFT that lead to a pronounced scattering and a modulation in DOS of the compact Fermi surface, which can vary from a minimum to maximum depending on the depth of SFT (see Supplemental Material Note \blue{I} for the minimum of LDOS at Fermi energy for SFT \#2). For the weak interband coupling and very low DOS ratio $\eta$, only an unshifted $\Delta_1$ is observed (green curve in Fig. \ref{fig:different interband coupling}(d)). For the medium interband coupling and medium DOS ratio $\eta$, two gaps move towards each with comparable coherence peak intensity (magenta curve in Fig. \ref{fig:different interband coupling}(e)). Eventually, for the strong interband coupling, the gaps have merged completely shown as red curve in Fig.~\ref{fig:different interband coupling}(f). We point out that the intraband scatterings play a role in broadening the coherence peaks [Fig. \ref{fig:influence of intraband couplings}]. The dashed lines in Fig. \ref{fig:different interband coupling}(a)-(f) mark the energy of coherence peaks of defect-free spectra (see top panel of Fig. 1(e)) for defect-free spectra).

Quasiparticle interference (QPI) patterns in 2 dimensions around the Fermi energy can be seen in the triangle in Fig. \ref{fig:SFT1}(c) and (d).
We point out that these QPIs are not due to superconducting states, as they do not change when superconductivity is suppressed by a magnetic field [Fig. 3(c)]. Therefore, these patterns arise from the scattering of unpaired quasiparticles around the Fermi energy. 
To quantify the QPI, we Fourier transformed the d$I$/d$U$ maps. Figure~\ref{fig:different interband coupling}(g)-(i) show the QPI patterns in reciprocal space. In SFT \#1 [Fig.~\ref{fig:different interband coupling}(g)], a wave vector of $q_1=0.13\times(2\pi/a)$ with six-fold symmetry can be seen. $a$ is the lattice constant of Pb. In SFT \#2 [Fig.~\ref{fig:different interband coupling}(h)], there are three different wave vectors with values of $q_1=0.13\times(2\pi/a)$, $q_2=0.40\times(2\pi/a)$ and $q_3=0.20\times(2\pi/a)$. In SFT \#3 [Fig.~\ref{fig:different interband coupling}(i)], a wave vector of $q_2=0.4\times(2\pi/a)$ and $q_3=0.2\times(2\pi/a)$ can be seen.

As the QPI pattern is sensed at the surface above a defect, the patterns are interpreted in the framework of the surface Brillouin zone, with the momentum perpendicular to the surface integrated, i.e. $k_{\perp}$ is undefined \cite{Hoffman2002Imaging,Liu2015Observation,Crommie1993Imaging}. 
To understand the QPI near the STF, we used first-principles calculations to obtain the projected surface DOS for the open and compact Fermi surface [Fig.~\ref{fig:different interband coupling}(j)] to figure out the nesting vectors in the (111) plane. Figure~\ref{fig:different interband coupling}(j) shows the calculated result. The observed scattering vectors (length and direction) can be found in the projected DOS to connect flat parts of the open and compact Fermi surfaces, as indicated by the cyan and yellow arrows. 

In SFT \#1, as discussed in [Fig.~\ref{fig:different interband coupling}(a) and (d)], the DOS ratio $\eta$ is very low in the triangle. This indicates that the LDOS on the compact Fermi surface is very small. Therefore, only the intraband scattering on the open Fermi surface is observed. The wave vector (cyan arrow) in Fig. \ref{fig:different interband coupling}(g) matches very well in Fig.~\ref{fig:different interband coupling}(j). However, for SFT \#2 [Fig.~\ref{fig:different interband coupling}(b) and (e)], the DOS ratio on the compact and open Fermi surfaces is comparable, so both interband scattering between the compact and open Fermi surfaces and intraband scattering on each Fermi surface are possible. Indeed, three different wave vectors are seen in Fig. \ref{fig:different interband coupling}(h). Interestingly, all the three wave vectors can be related the nesting vectors in Fig. \ref{fig:different interband coupling}(j), namely $q_1$ and $q_2$ (cyan arrows) corresponds to the intraband scattering of open Fermi surface and compact Fermi surface, respectively. And $q_3$ is the nesting of interband scattering (yellow arrow) in Fig.~\ref{fig:different interband coupling}(j). These results are consistent with our previous analysis of the superconducting spectra, i.e. that strong intra- and interband scattering is present in the triangle region.
Now we come to SFT \#3 with strong interband coupling on triangle. On the triangle of SFT \#3, the DOS ratio is very large [Fig. \ref{fig:different interband coupling}(c) and (f)]. This indicates the LDOS of compact Fermi surface is very large, while the LDOS of open Fermi surface is small. For this reason, we can expect to observe both inter- and intra-band scattering. However, the intraband scattering of open Fermi surface might not be seen due to the weak LDOS. The two wave vectors (cyan and yellow arrows) in Fig. \ref{fig:different interband coupling}(i) correspond to the intraband nesting of compact Fermi surface and interband nesting between these two Fermi surfaces in Fig. \ref{fig:different interband coupling}(j).

\subsection{Spatially resolved inter- and intraband couplings}

As demonstrated in the previous section, both QWS, i.e. intraband scattering, and interband couplings have a pronounced influence on the superconducting gaps. After discussing the general behavior in the two distinct regions (triangle and hexagon), we focus on the spatial dependence of intra- and interband scattering. 

\begin{figure*}[!htb]
    \centering
    \includegraphics[width=1\textwidth]{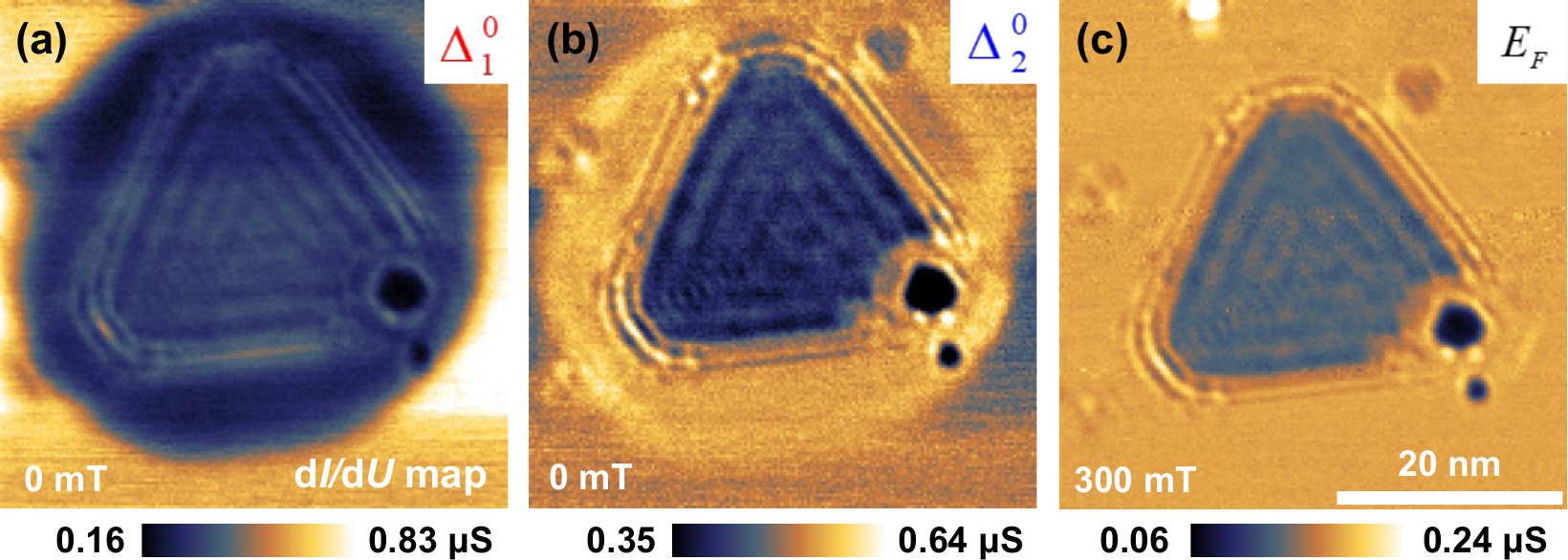}
    \caption{\textbf{SFT\#2 with medium interband coupling.} (a) d$I$/d$U$ map at $\Delta_1^0$. (b) d$I$/d$U$ map at $\Delta_2^0$. (c) d$I$/d$U$ map for normal states at Fermi energy measured at \SI{300}{\milli\tesla}. The tip was stabilized at $U=100$ mV and $I=1$ nA, with tip height z offset -120 pm for the d$I$/d$U$ spectroscopy. The lock-in amplifier has a modulation with amplitude of $U_{mod}=20$ $\mu$V and frequency of 3.402 kHz.}
    \label{fig:SFT2}
\end{figure*}

As STM allows fully spatial resolution, we mapped both intra- and interband coupling near defects. Figure~\ref{fig:SFT2} shows $dI/dU$-maps on SFT \#2 with an edge length of $\sim\SI{20}{\nano\meter}$. Here, the triangle region appears as a depression in the LDOS at both energies of the bulk coherence peaks, $\Delta_1^0$ [Fig.~\ref{fig:SFT2}(a)] and $\Delta_2^0$ [Fig.~\ref{fig:SFT2}(b)], due to a minimum in the normal DOS at the Fermi energy $E_F$ [Fig.~\ref{fig:SFT2}(c)] caused by the QWSs forming between this SFT and the surface (see also Fig. S1). Note that the $dI/dU$-map in Fig.~\ref{fig:SFT2}(c) was recorded at the Fermi energy in the normal state by applying a vertical magnetic field of $\SI{300}{\milli\tesla}$. We can clearly see that the decreased/increased $dI/dU$ signal in hexagonal region in the superconducting state shown in Fig.~\ref{fig:SFT2} (a)/(b) is absent in the normal conducting state Fig.~\ref{fig:SFT2}(c). This again highlights that the observed features are related to the two condensates and not to simple variations of the normal state DOS. We point out that the quasiparticle interference in the triangle region does not change with magnetic field, which originates from the nesting of two Fermi surfaces [Fig. \ref{fig:different interband coupling}].

On the SFT shown in Fig.~\ref{fig:SFT2}, we recorded $dI/dU$ spectra in a finely spaced grid and fitted them to the S\&W model DOS to obtain locally resolved information on the inter- and intraband scattering rates for each point and visualize them in a 2D map around the SFT. The results are presented in Fig.~\ref{fig: SFT2 eta and couplings} (see Supplemental Material Note \blue{V} for fitted $\sigma_i$). 
We can see that the defect locally induced variations of the DOS ratio $\eta$, as well as inter- and intraband scattering. Notably, $\eta$ has significantly enhanced in the hexagonal region [Fig. \ref{fig: SFT2 eta and couplings}(a)]. Interband scattering couples the gaps such that their sizes approach each other. Most importantly, interband scattering is enhanced in both the hexagonal in triangular regions, which is in the medium interband coupling regime [Fig. \ref{fig: SFT2 eta and couplings}(b)].  Intraband scattering is responsible for broadening the coherence peaks (see Appendices C). Intraband scattering on the open Fermi surface ($\Gamma_1$) is enhanced in the hexagonal and the triangular region [Fig.\ref{fig: SFT2 eta and couplings}(c)], while on the compact Fermi surface ($\Gamma_2$), it is only enhanced in the hexagonal region [Fig.\ref{fig: SFT2 eta and couplings}(d)]. For the case of strong interband coupling,  see SFT \#3 in Figs. \ref{fig:SFT3} and \ref{fig: SFT3 eta and couplings}.

These results illustrate, how scattering in two band superconductors couples the two order parameters and demonstrates that this coupling can be locally resolved and even engineered. Both extremes, i.e. of mostly decoupled condensates with two independent gaps and fully coupled condensates with only a single gap, can be realized in near proximity to each other in the same material.

\begin{figure}[!htb]
    \centering
    \includegraphics[width=\columnwidth]{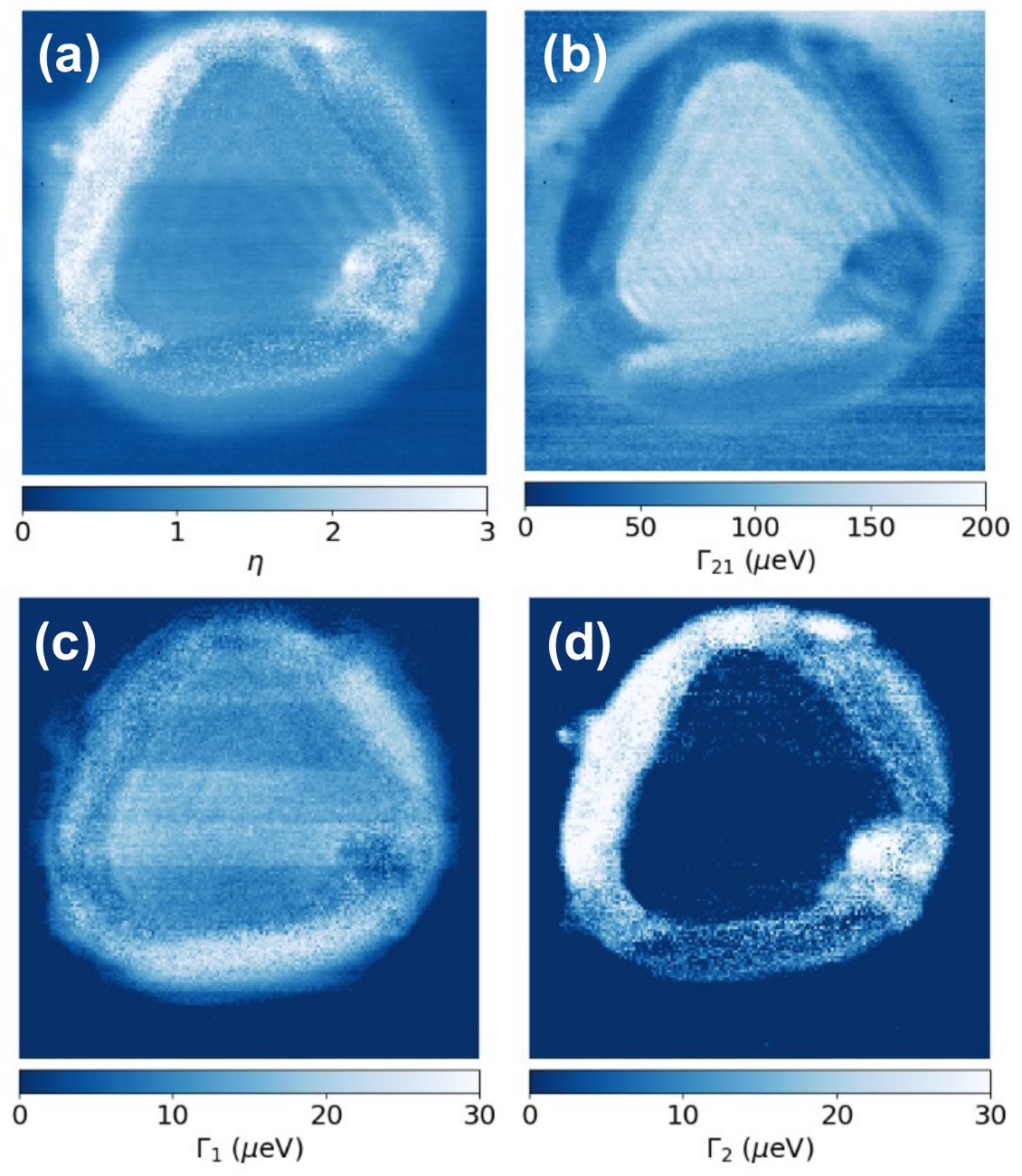}
    \caption{\textbf{Spatially resolved DOS ratio, inter- and intraband couplings (SFT\#2).} (a) Spatially resolved dos ratio $\eta$. (b) Spatially resolved interband coupling $\Gamma_{21}$. (c) Spatially resolved intraband coupling $\Gamma_{1}$. (d) Spatially resolved intraband coupling $\Gamma_{2}$.}
    \label{fig: SFT2 eta and couplings}
\end{figure}

\section{Discussion}

Our results show that a topological defect in form of a stacking fault locally alters the superconducting properties in a multiband system via drastic changes in the intra- and interband coupling in agreement with the expectation of a model put forward by Sung and Wong \cite{Sung-Wong1967}, where both inter- and intraband scattering processes are included. The intraband coupling is mostly responsible for lifetime broadening of the individual coherence peaks, while the interband coupling changes their relative energies. Eventually, strong interband coupling merges the gap magnitudes to a common value, making the density of states indistinguishable from the one of a single band system.
Intra- and interband scattering events can be understood on basis of the two Fermi surfaces. At a stacking fault, electrons of each band will be partially reflected staying on the same band, as the electronic dispersion relation at both sides of the stacking fault differs due to the Fm-3m symmetry of fcc, i.e. the Fermi surface only shows a threefold rotational symmetry and a stacking fault (stacking ABCABC$|$BCABC, where the missing 'A' layer creates the fault) represents a shift by about one‐third of the full lattice translation along the close‐packed (111) plane. Moreover, the stacking fault also induces scattering between the two Fermi surfaces, as the momentum of the electron perpendicular to the stacking fault is not conserved.  Fits of the spectra to the Sung and Wong model show an enhancement of the interband scattering 
$\Gamma_{21}$ on top of the whole stacking fault of the SFT i.e., triangular region in Fig. \ref{fig: SFT2 eta and couplings}(b), corresponding to the medium interband coupling shown in Fig. \ref{fig:different interband coupling}. This enhanced interband scattering extends out even beyond the SFT most likely due to electronic states with a finite momentum parallel to the surface. 
Note that the two scattering rates need to conserve the total number of electrons on each band, i.e. they are proportional to the DOS of each band (see Appendices A Methods).
Intraband scattering, however, may independently vary on each band as the reflection probability of Bloch states on the two bands by a stacking fault depends on the dispersion of the respective band. This is also reflected by the difference in the pattern of Figs. \ref{fig: SFT2 eta and couplings} (c) and (d). Due to the more symmetric shape of the compact Fermi surface, reflection at stacking faults is less expressed in comparison to the more asymmetric structure of the open Fermi surface. This is reflected in the generally higher values for intraband scattering $\Gamma_1$ in the triangle region. Similarly, intraband scattering is visible in the hexagonal region showing standing waves corresponding to the compact Fermi surface. 

The engineering of interband coupling is also of more fundamental interest. The pristine Pb crystal possesses two nearly decoupled condensates that can be coupled through interband scattering at defects. Especially, if the superconducting phase of the two bands laterally differ, e.g. by introducing a magnetic field, this engineering can lead to frustration of excitations such as vortices of solitonic modes. The control over the interband coupling in a two-band superconductor may allow to test for several predicted quantum effects, such as solitons \cite{Tanaka2002, Yerin2011}, vortices with fractional flux \cite{Babaev2002}, non-Abrikosov vortices \cite{Cho2006}, topological knots \cite{Cho2006},
or the Leggett mode \cite{leggett_number-phase_1966}. For example, the Leggett mode or soliton energy is a function of interband coupling, and a shift of these modes as a function of distance from an SFT would verify the nature of the modes.

\section{Acknowledgements}
We are grateful for discussions with Roland Willa, Horst Hahn, Marcel Rost and Reinhard Schneider. The authors acknowledge funding by the Deutsche Forschungsgemeinschaft (DFG) under the CRC TRR 288 - 422213477 grant “ElastoQMat”, projects B06 and B01, and support by JSPS KAKENHI under grant number 17K19023.

Experiments have been carried out by T.G., Q.L., A.H., R.N. and supervised by W.W.. The experimental data have been evaluated by T.G., Q.L. and O.K.. DFT calculations have been performed by R.H.. Theory was performed by Q.L., T.G., J.S. and W.W.. The manuscript was written by Q.L., T.G., and W.W. with input from all authors. The sample was supplied by R.N. and T.K.Y. 

The authors declare that they have no competing interests.

\section{Data availability}
All data needed to evaluate the conclusions in the paper are presented in the paper and in the Supplemental Materials.

\section{Appendices}
\subsection{Methods}
\paragraph{Experimental details:}
All experiments were performed using a home-built scanning tunneling microscope with dilution refrigeration at $45\,\mathrm{mK}$ \cite{balashov_compact_2018}. The bias voltage $U$ was applied between the sample and the common machine ground so that a positive bias voltage probes the unoccupied states of the sample. The STM chamber was kept at a base pressure of $1\times 10^{-10}\,\mathrm{mbar}$. The hat-shaped single crystal Pb(111) (miscut angle: $\pm 0.1^\circ$,  purity: $99.999\%$) has been purchased from MaTecK GmbH. At a base pressure of $1\times 10^{-10}\,\mathrm{mbar}$ the Pb crystal was prepared in cycles of hot sputtering with Ar$^+$ ions of $3\,\mathrm{k}e\mathrm{V}$ at moderate temperatures ($\sim 30-60\,\SI{}{\celsius}$) and subsequent annealing at $190^{\circ}$C. Following that, it was pre-cooled on a $\SI{78}{\kelvin}$ stage and then directly transferred into the STM in-situ, where it was cooled down to the base temperature in zero field. A tungsten tip was prepared by high-temperature flashing and soft dipping into a Au(111) surface. The differential conductance was measured using a Lock-in amplifier at a frequency of $3.0-3.8\,\mathrm{kHz}$. 

\paragraph{Sung-Wong model:}
To extract inter- and intraband scattering induced by SFTs, the S\&W model \cite{Sung-Wong1967} was used to fit tunneling spectroscopy of superconducting gaps. First, tunneling spectroscopy is the sum of two BCS from tunneling DOS with temperature broadening.

\begin{align}
    \frac{\mathrm{d}I}{\mathrm{d}U}(eU) =  \sum\limits_{i = 1{,}2} \sigma_i \operatorname{Re}(\frac{|u_i|}{\sqrt{u_i^2-1}})\ast f^\prime\label{Eq: DOS}.
\end{align}

$\sigma_i$ denotes the band-specific differential conductance with $\sigma_i=W_i N_i(E_F)$ \cite{Noat2010signatures, Noat2010TwoEnergyGaps, noat_quasiparticle_2015}. Hereinafter $i=1, 2$ denotes the band index. $W_i$ is the momentum averaged tunneling probability. $N_i(E_F)$ is the DOS at Fermi energy.  Re is the real part of the function. $f^\prime$ is the derivative of the Fermi-Dirac distribution with respect to energy at temperature $T$. $\ast$ is the convolution due to the temperature-induced broadening. 

Second, $u_i$ are determined by the coupled equations that are self-consistent \cite{Sung-Wong1967, Schopohl-Scarnberg1977, Iavarone2002Two-band},

\begin{align}
   u_1 \Delta_1^0=|E|+i\Gamma_1 +i\Gamma_{12}\frac{u_2-u_1}{\sqrt{u_2^2-1}}\label{Eq: delta1},\\
   u_2 \Delta_2^0=|E|+i\Gamma_2+i\Gamma_{21}\frac{u_1-u_2}{\sqrt{u_1^2-1}}\label{Eq: delta2}.
\end{align}

$\Gamma_\lambda$ is the intraband coupling, and $\Gamma_{ij}$ is the interband coupling with scattering from band $i$ to $j$. In addition, $\Gamma_{12}$ and $\Gamma_{21}$ are related by the equilibrium condition.

\begin{align}
    \frac{\Gamma_{12}}{\Gamma_{21}}=\frac{N_2(E_F)}{N_1(E_F)}\label{Eq: gamma}.
\end{align}

$\Delta_\lambda^0$ is the intrinsic order parameter, which is caused by the phonon-mediated attractive interaction \cite{Karakozov2010}. We argue that there is only intraband pairing in Pb, as interband pairing results in pair density wave, which is not observed in our measurements. In the defect-free area, the scatterings are very small. The ratio $\eta_0=\frac{N_{02}(E_F)}{N_{01}(E_F)}=0.46$ can be obtained from the first-principles calculations with $N_{02}(E_F)=0.162$ $(1/eV)$ for the compact Fermi surface and $N_{01}(E_F)=0.335$ $(1/eV)$ for the open Fermi surface. By this we obtained the tunneling probability $W_i$ for each band in our measurements. These tunneling probabilities are then used in the S\&W model fitting for SFTs. We fitted the defect-free spectroscopy and obtained $\Delta_i^0$. 
We treat $\Delta_i^0$ as constant when fitting data in the vicinity of SFTs. Then, the gap changes are induced by superconducting proximity due to SFT scatterings. We point out that Pb is a conventional two-band superconductor with same sign (same phase) of the gap on the different Fermi surface sheets. 
As a result of impurity scattering, the gaps become energy-dependent and complex in Eqs. \ref{Eq: delta1} and \ref{Eq: delta2}. We note that the intra- and interband couplings are homogeneous where there are no defects or impurities. However, the intra- and interband scatterings are spatially dependent around defects or impurities.

\paragraph{First-principles calculations:}
Density functional calculations of the electronic structure of Pb were carried out in the framework of the mixed-basis pseudopotential method \cite{Elsasser-Relativistic1990, Meyer-Ab-initio1997}. The electron-ion interaction was represented by norm-conserving relativistic pseudopotentials \cite{Vanderbilt-optimally1985}. Spin-orbit coupling was incorporated within the pseudopotential scheme via Kleinman's formulation \cite{Kleinman-Relativistic1980} and was consistently taken into account in the charge self-consistency cycle using a spinor representation of the wave functions. Further details of the spin-orbit coupling implementation within the mixed-basis pseudopotential method can be found in a previous publication \cite{Heid-effect-2010}. For higher accuracy, 5$d$ semicore states were included in the valence space. The deep $d$ potential is efficiently treated by the mixed-basis approach, where valence states are expanded in a combination of plane waves and local functions.  Here, local functions of $d$ type at the Pb sites were combined with plane waves up to a kinetic energy of 20~Ry. Brillouin-zone integration was performed by sampling a 32$\times$32$\times$32 $k$-point mesh (corresponding to 2992 $k$ points in the irreducible part of the Brillouin zone) in conjunction with a Gaussian broadening of 0.2~eV. The exchange-correlation functional was represented by the local-density approximation in the parameterization of Perdew-Wang \cite{Perdew-accurate-1992}.
In this way, we obtained the DOS at the Fermi energy for the open ($N_{01}(E_F)$) and compact ($N_{02}(E_F)$) Fermi surfaces via first-principles calculations. For the 2D-projected DOS of both Fermi surfaces, these results were obtained by first calculating band energies on a dense $100\times100\times100$ $k$ mesh. The energies were then linearly interpolated to perform the projection using 800 points along $k_z$ and a Gaussian broadening of 0.1 eV.

\subsection{Stacking fault tetrahedron on Pb(111)}
\begin{figure}[!htb]
    \centering
    \includegraphics[width=1\columnwidth]{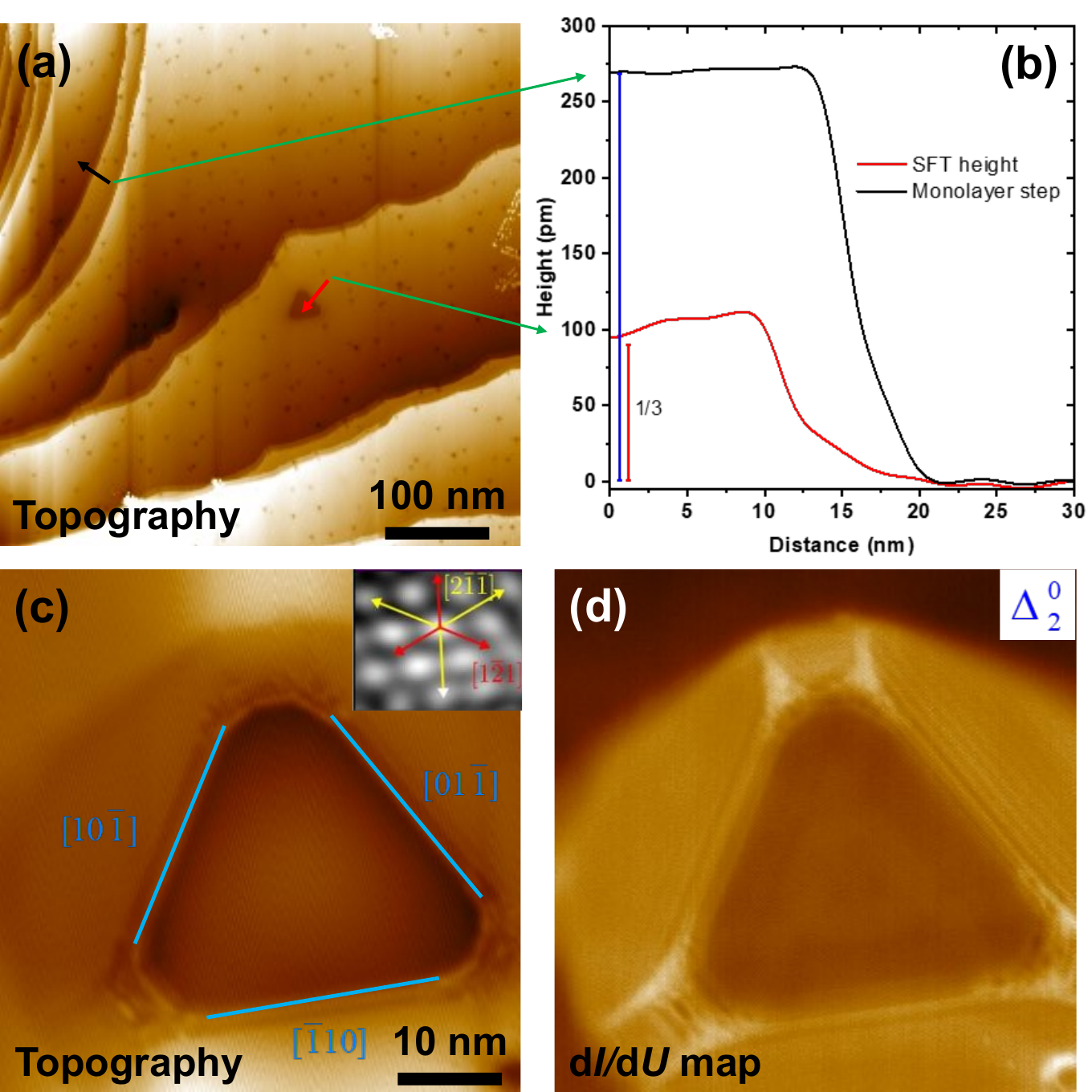}
    \caption{\textbf{Stacking fault tetrahedron on Pb(111) (SFT \#4)}. (a) An SFT with a side length of $\sim$30 nm on the Pb(111) surface. The tip was stabilized at $U=100$ mV and $I=50$ pA. (b) The distance between the SFT and the surface is 95 pm, which is equal to 1/3 of the monolayer step height. 
    Note that the next lower terrace is set to zero as reference. This is a fingerprint of an SFT on a (111) surface. (c) Zoom-in topography shows that the directions along which the edges of the triangle are oriented are $<110>$, which is identified by the atomic resolution in the inset. The edges of the triangle are actually the stair rods to two intersected {111} surfaces. (d) The d$I$/d$U$ map at the energy of $\Delta_2^0$ shows the enhanced superconducting density of states outside the triangle. The tip was stabilized at $U=1.40$ mV and $I=100$ pA. The lock-in amplifier had modulation of 50 $\mu$V and frequency of 3.751 kHz. }
    \label{fig:SFT4 on Pb}
\end{figure}

Stacking fault tetrahedrons (SFTs) are low-energy defects in face-centered cubic (fcc) crystals that have been extensively studied, e.g. in Au \cite{silcox_direct_1959, Cotterill_1961, Yokota_1967, Matsukawa_2008, Wang_atomic_2013, schouteden_electronically_2016}, Ag \cite{Wolf1991, Hardy_2006, schouteden_lateral_2012, Yu_removal_2013}, Cu \cite{schaublin_irradiation-induced_2005}, and Al \cite{Guan_stacking_2004}. However, they are less reported in Pb(111) \cite{Bonsignori_1968}.  Figure \ref{fig:SFT4 on Pb}(a) shows a topography with terraces and monolayer steps. On the terraces, there are many small Ar-induced vacancies and Ar bubbles buried beneath the surface. A screw dislocation can also be seen. In addition, there is a dark triangle with a lateral length of about 30 nm on the surface [Fig.~\ref{fig:SFT4 on Pb}(a)] that is a stacking fault tetrahedron on Pb(111). First, the step height on the surface above the SFT is approximately 1/3 of a monolayer [Fig.~\ref{fig:SFT4 on Pb}(b)] in agreement with the expected Burgers vector and similar results on SFTs reported by Wolf and Ibach \cite{Wolf1991}. Second, from the atomic resolution of the surface, we identify the edges of the triangle that follow the $\langle 110\rangle$ directions on the (111) surface [Fig.~\ref{fig:SFT4 on Pb}(c)], which are the directions of the expected stair-rod dislocations of the SFT \cite{Ohmori1999, Osetsky2006}. Due to crystallography, all SFTs in our sample have a similar shape and orientation. In fact, we observed more than 20 SFTs and the orientation of the triangle, i.e. the orientation of the SFT, is the same in all cases. The side lengths of the SFTs range from 20 to more than 100 nm. In contrast to normal conducting fcc metals, Pb allows to explore the influence of SFTs on superconductivity. Figure~\ref{fig:SFT4 on Pb}(d) shows the corresponding d$I$/d$U$ map at $\Delta_2^0$, which shows an enhanced superconducting density of states outside the triangle similar to the SFTs shown in Figs. \ref{fig:SFT1}(d) and \ref{fig:SFT2}(b).

\subsection{Influence of intraband coupling}
\begin{figure}[!htb]
    \centering
    \includegraphics[width=1\columnwidth]{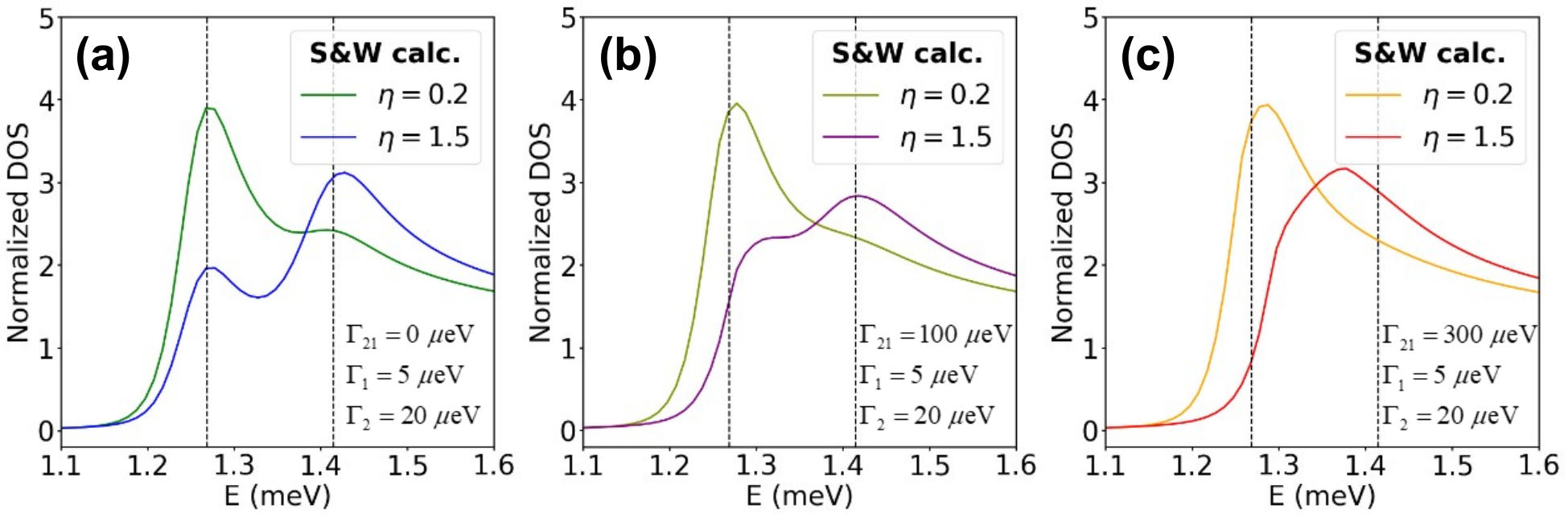}
    \caption{\textbf{Influence of intraband couplings.} The broadening effect of intraband couplings decreases the intensities of coherence peaks and rounds the gap edges.}
    \label{fig:influence of intraband couplings}
\end{figure}

The intraband couplings can also modify the quasiparticle spectroscopy of superconducting gaps. Figure \ref{fig:influence of intraband couplings} demonstrates the influence of intraband couplings. We can see that the intensities of coherence peaks decrease and the gap edges are not sharp any more compared with Figs. 2(a)-(c), which match the experimental results better. It indicates that intraband couplings should also be taken into account. 

\subsection{SFT with strong interband coupling}

\begin{figure}[!htb]
    \centering
    \includegraphics[width=1\columnwidth]{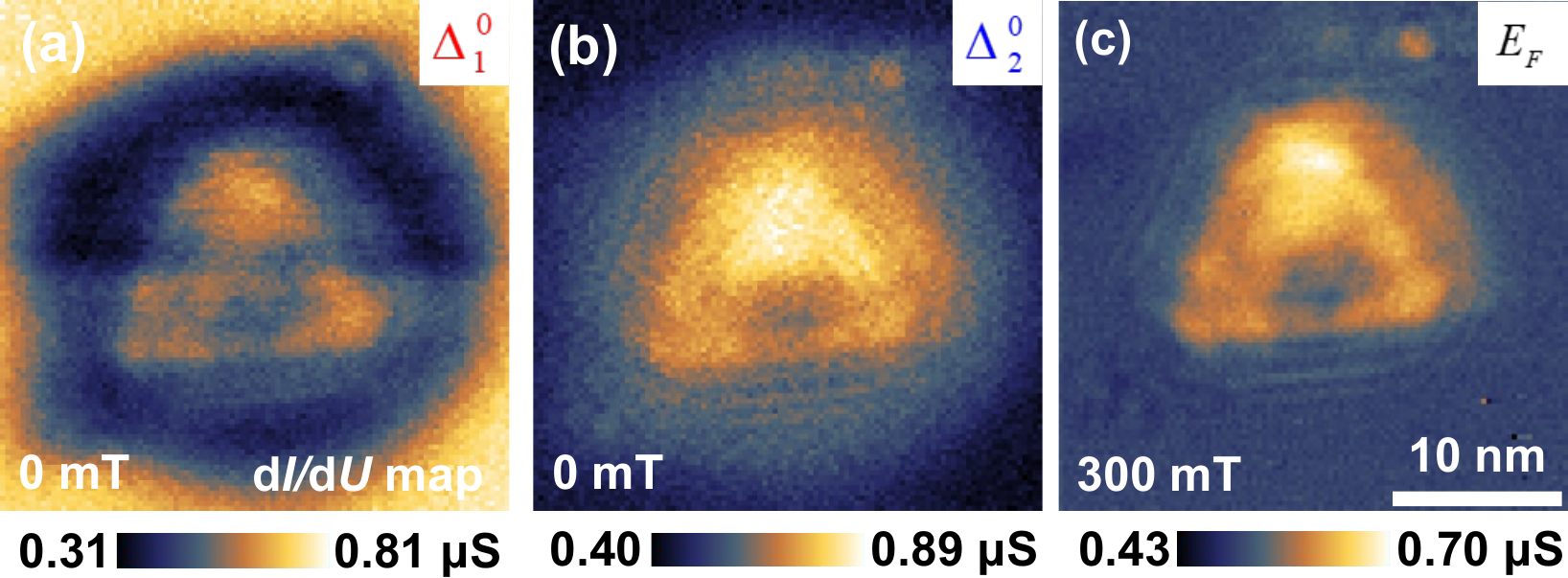}
    \caption{\textbf{SFT \#3 with strong interband coupling.} (a) d$I$/d$U$ map at energy $\Delta_1^0$. (b) d$I$/d$U$ map at energy $\Delta_2^0$. (c) d$I$/d$U$ map at energy $E_F$ with magnetic field 300 mT. In (a) and (b), the tip was stabilized at $U=100$ mV and $I=1$ nA with tip height z offset -120 pm, and lock-in modulation with amplitude of $U_{mod}=20$ $\mu$V and frequency of 3.402 kHz for the d$I$/d$U$ spectroscopy. In (c), the tip was stabilized at $U=100$ mV and $I=1$ nA, with tip height z offset -160 pm, and lock-in modulation with amplitude of $U_{mod}=35$ $\mu$V and frequency of 3.402 kHz for the d$I$/d$U$ spectroscopy.}
    \label{fig:SFT3}
\end{figure}

SFT \#3 exhbits strong interband coupling, which has a side length about 16 nm. In SFT \#3, a similar contrast in hexagon region for d$I$/d$U$ maps at energy $\Delta_1^0$ Fig.~\ref{fig:SFT3}(a) and $\Delta_2^0$ Fig.~\ref{fig:SFT3}(b). SFT \#3 has an enhanced DOS in the triangle region Fig.~\ref{fig:SFT3}(c), which is distinct from SFT \#2 Fig.~\ref{fig:SFT2}(c). This is due to its different depth causing an enhanced LDOS by the QWS. 

\begin{figure}[!t]
    \centering
    \includegraphics[width=1\columnwidth]{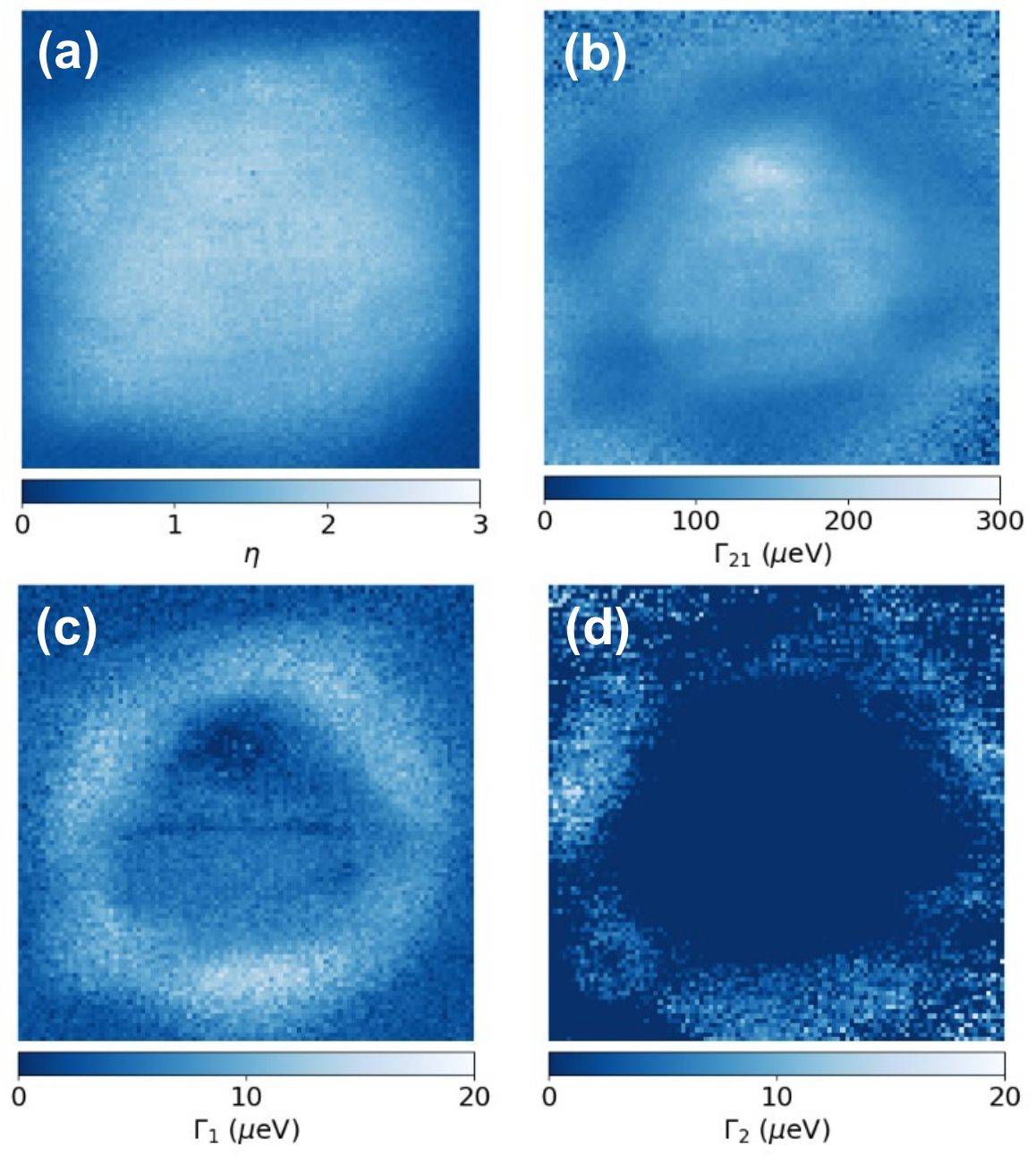}
    \caption{\textbf{The spatially resolved DOS ratio, inter- and intraband couplings for SFT \#3.} (a) Spatially resolved DOS ratio $\eta$. (b) The interband coupling $\Gamma_{21}$. (c) The intraband coupling $\Gamma_{1}$. (d) The intraband coupling $\Gamma_{2}$.}
    \label{fig: SFT3 eta and couplings}
\end{figure}

The fits of DOS ratio, inter- and intraband coupling are shown in Fig.~\ref{fig: SFT3 eta and couplings}. In Fig. \ref{fig: SFT3 eta and couplings}(a), the DOS ratio $\eta$ in the triangle is even higher than that in the hexagon, which in contrast with the lower $\eta$ in the triangle than that in the hexagon in Fig.~\ref{fig: SFT2 eta and couplings}. which have patterns similar to those in Fig. \ref{fig: SFT2 eta and couplings}.

\FloatBarrier

\bibliography{main.bib}


\end{document}